\documentclass[fleqn,twoside]{article}
\usepackage{espcrc2}

\usepackage{graphicx}
\usepackage[figuresright]{rotating}

\def\beq{\begin{equation}} 
\def\beq#1{\begin{equation} \label{#1}}
\def\eeq{\end{equation}}
\newcommand{\bea}{\begin{eqnarray}}
\newcommand{\eea}{\end{eqnarray}}
\newcommand{\ber}{\begin{eqnarray}}
\newcommand{\eer}{\end{eqnarray}}
\newcommand{\bers}{\begin{eqnarray*}}
\newcommand{\eers}{\end{eqnarray*}}

\newcommand{\AmS}{{\protect\the\textfont2
  A\kern-.1667em\lower.5ex\hbox{M}\kern-.125emS}}

\title{PUZZLES IN HYPERON, CHARM AND BEAUTY PHYSICS}

\author{Harry J. Lipkin\address{Department of Particle Physics, 
        Weizmann Institute of Science,  
        Rehovot, Israel }  
	 \address{School of Physics and Astronomy,
Raymond and Beverly Sackler Faculty of Exact Sciences,
        Tel Aviv University, Tel Aviv, Israel }%
\thanks  {Supported
in part by a grant from the United States-Israel
Binational Science Foundation (BSF), Jerusalem, Israel,
by the Basic Research Foundation administered by the Israel Academy of 
Sciences and Humanities} 
      \address{High Energy Physics Division, Argonne National Laboratory,\\ 
Argonne, IL 60439-4815, USA}  
\thanks  {Supported
in part by the U.S. Department
of Energy, Division of High Energy Physics,
Contract W-31-109-ENG-38.}}

\begin{document}

\begin{abstract}
Puzzles awaiting better experiments and better theory include: (1) the
contradiction between good and bad SU(3) baryon wave functions in fitting
Cabibbo theory for hyperon decays, strangeness suppression in the sea and the
violation of the Gottfried Sum rule - no model fits all; (2) Anomalously
enhanced Cabibbo-suppressed $D^+ \rightarrow K^{*+}(s\bar d)$ decays; (3)
anomalously  enhanced and suppressed $B \rightarrow \eta' X$ decays; (4) the
OZI rule in weak decays;   (5) Vector dominance ($W \rightarrow \pi, \rho, a_1,
D_s, D^*_s $) in weak decays (6) Puzzles in doubly-cabibbo-suppressed charm
decays.(7) Problems in obtaining $\Lambda$ spin structure from polarization
measurements of produced $\Lambda$'s.
\vspace{1pc}
\end{abstract}

% typeset front matter (including abstract)
\maketitle

\section{PREAMBLE - CHALLENGING THE CONVENTIONAL WISDOM}

\subsection{The war for quarks}

In the introduction to this session we were reminded of the war to
convince the Physics  Community that quarks are real.

We won that war.

\subsection{The battle for science and education}

My next war is to convince the University Community and University Departments
of Education that Physics is real and not socially constructed. High
school teachers should  know and respect science, not only politically correct
education ideology. It is not an easy battle. I participated in one round at the Peter Wall Institute for Advanced Studies here at UBC in
May.

More about this war and similar approaches to understanding the structure of
matter and understanding how to teach children can be found in my millennium
essay "The Structure of Matter - like science, teaching should be the result of
independent ideas converging"\cite {natmil}.  
                         
\subsection{Following up experiments that challenge conventional wisdom}

Just as we learned new things by following up the experimental evidence for
quarks even though there was no theory, we can do the same today with new
experimental puzzles. Look for new experiments that can bring new light and
insight.
 
For example, the recent CP asymmetry in $B \rightarrow \Phi K_S$ decay,
considered a hint of non-standard CP violation\cite{hiller}, suggests a search
for a direct CP-violation in the charged B decays, $B^{\pm} \rightarrow  \Phi
K^{\pm}$, $B^{\pm} \rightarrow \Phi K^*$, where $K^*$ denotes any $K^*$
resonance, and $B^{\pm} \rightarrow \Phi K_S \pi^{\pm}$. The $\Phi K_S
\pi^{\pm} $ analysis can use transversity to distinguish between  different new
physics models  predicting different behaviors for a charge asymmetry as a
function of the transversity and of the mass of the $K \pi$ system.  A
new-physics CP-violating contribution to $B^{\pm} \rightarrow  \Phi K^{\pm}$
and  $B \rightarrow \Phi K^*$ could arise from    the  $b \rightarrow s $
transition induced by gluino exchange\cite{kagneub} applied to the radiative
decay $B \rightarrow X_s \gamma$, with the photon materialized into a $\Phi$.

\section{PUZZLES IN HYPERON PHYSICS}
 
\subsection{Good and bad SU(3) baryons}
 
So far no model for the flavor structure of the
proton  is simultaneously consistent with three established
experimental results\cite{nucabib}:
\begin{enumerate}
\item
    $SU(3)$  Cabibbo theory  good for weak semileptonic baryon decays
\item
$SU(3)$ badly  broken by the suppression
of the strange component in the sea of $q \bar q$ pairs in the nucleon
 by a factor of two\cite{CCFR}.
\item
The observed violation of the Gottfried sum rule 
requires  a positively charged sea ($u\bar d$).
$SU(3)$ then requires a sea with net strangeness $(u\bar s$). 
\end{enumerate}

Explaining the observed violation of the Gottfried sum rule while keeping the
good results of the Cabibbo theory requires unobserved  net
strangeness in the nucleon sea. Two possible directions for avoiding this
conflict are\cite{nucabib}:

\begin{enumerate}
\item
Quantitative analysis of how much violation of Cabibbo theory is allowed by the
real hyperon decay data with real errors. 

\item
Small component of 
``valence like" strange quarks with large values of $x$ in the proton; 
This can be checked by
better measurements of the $x$-dependence of the strangeness in the proton. 

\end{enumerate}

\subsection{Is the strange quark a heavy quark?}

 Most nontrivial strange hadron states mix SU(3) and satisfy (scb) heavy quark
symmetry with no flavor mixing;  e.g. $\phi, \psi, \Upsilon$.  The strange
axial meson mixing is still open. HQET\cite{Iswis}
couples the light quark spin to the orbital $L=1$ to get two doublets with j=1/2
and j=3/2.   SU(3) has  triplet and singlet.spin states in the octets
with the $a_1$ and $b_1$. Which is it?

\subsection{The spin structure of the $\Lambda$}

In the constituent quark model the  strange quark carries the entire
spin of the baryon. But in the picture where strange quarks carry some of the
spin of the proton, the nonstrange quarks carry some of the spin of the 
$\Lambda$. Possible experimental tests of this description were 
discussed\cite{Ma}.  

\section{{A VECTOR DOMINANCE MODEL}}

\subsection{{Vector dominance universality}}

The large branching ratios observed\cite{PDG}  for the appearance of the
$a_1(1260)^{\pm}$ in all quasi-two-body decays $D \rightarrow  a_1(1260)^{\pm}
X$ and  $B \rightarrow  a_1(1260)^{\pm}  X$  are comparable to  those observed
for  $\pi^{\pm} X$ and  $\rho^{\pm} X$   and contrast sharply with the much
smaller branching ratios observed to  $a_2 X$,  $b_1 X$,  and  $a_1^o X$.  

 All 24 $B$ decays of the form $ B \rightarrow  \bar D
W^+\rightarrow  \bar D M^+ $, where $M$ can denote  $a_1, \rho,  \pi,
{\ell}^+\nu_{\ell}, D_s, D^*_s $, are dominant with branching ratios above 
$0.3\%$. Other $B$-decay modes have upper  limits in the $10^{-4}$ ball park,
including 
the absence with significant upper limits of  neutral decays $B^o \rightarrow
\bar D^o M^o $ which are coupled by strong final state interactions to $B^o
\rightarrow D^- M^+ $.

These experimental systematics suggested a 
``vector-dominance" model\cite{vecdom} where the initial hadron state $i$
decays to a final state $f$ by emitting a $W^{\pm}$ which then hadronizes into
an $a_1^+$, $\rho^+$ or $\pi^+$, along  with a universality relation,
\beq{veca}
{R[ifa]} \equiv {{BR[ i \rightarrow  f  a_1(1260)^+]}\over{BR[i
\rightarrow  f \rho^+] }}\approx \left|{{W^+  \rightarrow  a_1^+ }\over{W^+ 
\rightarrow   \rho^+ }}\right|^2  
\eeq
\beq{vecpi}
{R[if\pi]} \equiv {{BR[ i \rightarrow  f \pi^+]}\over{BR[i
\rightarrow  f \rho^+] }}\approx \left|{{W^+  \rightarrow  \pi^+ }\over{W^+ 
\rightarrow  \rho^+ }}\right|^2   
\eeq

These have been shown to hold experimentally\cite{vecdom} for all states $i$ 
and $f$ with corrections for phase space.  The $a_1$ data have large errors.
But the experimental ratios  $R[ifa]$ are all consistent with 0.7,  and more
than order of magnitude higher than other upper limits.
That such widely different decays should agree so well is impressive  and
suggests further investigation. e.g. reducing the experimental
errors and looking for more decay modes like 
$D^+_s \rightarrow \phi a_1$,
$D^+  \rightarrow K^{*0} a_1$ and
$D^0  \rightarrow K^{*-} a_1$.

\subsection {Vector-Dominance Decays of the $B_c$}

The $B_c$ meson is identified against a large combinatorial background by decay
modes including a $J/\psi$. 
Vector dominance decay modes including the $J/\psi$ are expected to have 
relatively large branching ratios\cite{vecdom}. These include:
$J/\psi \rho^+ $, $J/\psi a_1^+ $,  $J/\psi \pi^+ $, $J/\psi D^*_s$,
$J/\psi D_{s1A}$, and $J/\psi D_s $. 
The corresponding modes with a $\psi'$ instead of a $J/\psi$ are expected to 
have comparable branching ratios.

\section{{PUZZLES IN CHARM DECAYS}}

\subsection{{Singly-Suppressed Charm Decays}}

Two Cabibbo suppressed $D^+$ decay modes have anomalously high branching ratios
which are not simply explained by any model\cite{nuclolip}.
\beq{nudnik01}  
BR[D^+ \rightarrow  K^*(892)^+\bar K^o] = 3.2 \pm 1.5\%
\eeq
\beq{nudnik02}  
BR[D^+ \rightarrow  K^*(892)^+\bar K^*(892)^o] = 2.6 \pm 1.1\%
\eeq
These show no Cabibbo
suppression in comparison with  corresponding Cabibbo allowed decays
whose dominant tree diagrams  differ only in the weak vertices $c \rightarrow W^+ + s
\rightarrow \rho^+ + s$ and $c \rightarrow W^+ + s
\rightarrow K^*(892)^+ + s$ from  corresponding allowed decay diagrams
and have the same hadronization of the strange quark
$s$ and spectator $\bar d$.
\beq{nudni03}  
BR[D^+ \rightarrow  \rho^+\bar K^o] = 6.6 \pm 2.5\%
\eeq 
\beq{nudnik04}  
BR[D^+ \rightarrow  \rho^+\bar K^*(892)^o] = 2.1 \pm 1.3\%
\eeq 
No simple diagram can contribute to the anomalously 
enhanced decays 
without also enhancing one of the following others which show the
expected Cabibbo suppression
\beq{nudnik05}  
BR[D^+ \rightarrow  K^+\bar K^*(892)^o] = 0.42 \pm 0.05\%
\eeq 
\beq{nudnik06}  
BR[D^o \rightarrow  K^*(892)^+K^-] = 0.35 \pm 0.08\%
\eeq 
\beq{nudnik07}  
BR[D^o \rightarrow  K^*(892)^-K^+] = 0.18 \pm 0.01\%
\eeq 
\beq{nudnik08} 
BR[D^o \rightarrow  K^*(892)^o\bar K^o] < 0.08 \%
\eeq 
\beq{nudnik09}  
BR[D^o \rightarrow \bar K^*(892)^oK^o] < 0.16 \%
\eeq 
\beq{nudnik10}
BR[D^o \rightarrow  K^*(892)^o\bar K^*(892)^o] = 0.14 \pm 0.05\%
\eeq 
A new physics explanation may be needed if the anomalously large branching
ratios are confirmed with smaller errors. Present data show\cite{PDG}
\bea
BR[D^+ \rightarrow  K^*(892)^+\bar K^*(892)^o]  + \nonumber \\
+ BR[D^+ \rightarrow  K^*(892)^+\bar K^o] = 5.8  \pm 1.9\%
\eea
This is still large even at two standard deviations. 

\subsection {SU(3) Relations between Cabibbo-Favored and Doubly-Cabibbo Suppressed
$D$ decays}.

The SU(3) transformation\cite{Lipkin1997} $d \leftrightarrow s $ relates Cabibbo-favored
$\leftrightarrow $ doubly-Cabibbo-suppressed charm decays\cite{plipzh}. 
\bea     
d \leftrightarrow s  ; ~ ~ ~    K^+ \leftrightarrow \pi^+  ; ~ ~ ~
  K^- \leftrightarrow \pi^-  ; \nonumber \\   
    D^+ \leftrightarrow D_s;  ~ ~ ~ 
 D^o \leftrightarrow D^o;  ~ ~ ~ 
   K^+ \pi^- \leftrightarrow  K^- \pi^+         
\eea

If strong interaction final state interactions
conserve SU(3) the only SU(3) breaking occurs in the CKM matrix elements.

\subsubsection{Relations between $D^o$ branching ratios}

Two simple easily tested.SU(3) symmetry relations involving 
no phases and only
branching ratios of decay modes all expected to be comparable to the observed
DCSD $D^o \rightarrow K^+ \pi^-$ are
\bea  
 tan^4 \theta_c  = {{BR( D^o \rightarrow  K^+ \pi^-)}\over{BR(D^o \rightarrow  
K^-\pi^+) }} = \nonumber \\ 
={{BR[ D^o \rightarrow  K^*(892)^+\rho^-]}\over{BR[D^o \rightarrow  
K^*(892)^-\rho^+]}} 
\eea
A similar relation 
\beq{nudnikb} 
tan^4 \theta_c =  
 {{BR[ D^o \rightarrow  K^+  a_1(1260)^-]}\over{BR[D^o \rightarrow  
K^- a_1(1260)^+] }} 
\eeq
may have a different type of SU(3) breaking. A weak vector dominance form factor
can enhance 
\beq{nudnikc}
D^o(c\bar u)
\rightarrow  (s \bar u\rightarrow K^{-} )_S \cdot 
(u\bar d\rightarrow  a_1^+ )_W
\rightarrow  K^{-} a_1^+ 
\eeq
where the subscripts S and W  denote strong and weak form factors.
 The largest 
SU(3) breaking may well arise here from the difference between a weak 
pointlike $a_1$ form factor and the strong  hadronic $a_1$ form factor which is 
the overlap between a nodeless s-wave meson  and a p-wave meson with a node.
This suppression  of the  $a_1$  hadronic form factor should suppress 
\beq{nudnikd} 
D^o(c\bar u)  \rightarrow  (d \bar u \rightarrow a_1^-)_S \cdot 
(u\bar s \rightarrow K^{+} )_W \rightarrow  a_1^-  K^{+ }  
\eeq
In this case the SU(3) relation between the ratios of two Cabibbo-favored 
decays to two
Cabibbo-suppressed decays involving the $\pi$ and $a_1$ can be expected to be
strongly broken and replaced by the  inequality
\bea
{{BR[D^o \rightarrow  
K^- a_1^+]}\over{BR(D^o \rightarrow  
K^-\pi^+) }} =  
{{7.3 \pm 1.1 \%}\over{ 
 3.83 \pm 0.09\%}} \gg \\ \nonumber
 \gg {{BR[ D^o \rightarrow  K^+  a_1^-]}\over{
BR( D^o \rightarrow  K^+ \pi^-) }} = 
{{BR[ D^o \rightarrow  K^+  a_1^-]}\over{
1.48\pm 0.21\times 10^{-4} }}
\eea 

 Two aspects of this relation suggest interesting implications of any
 symmetry breaking:

 (1)  Experimental tests of the magnitude of SU(3) breaking will be relevant
in the interpretation of information about the CKM matrix and the unitarity
triangle obtained from standard model analyses of weak decays which assume 
SU(3) symmetry. 

  (2) In the standard model the Cabibbo-favored and doubly-suppressed charm
decays are proportional to the same combinations of CKM matrix elements and
no direct CP violation can be observed. Thus any evidence for new physics that
can introduce a CP-violating phase between these to amplitudes deserves serious 
consideration\cite{plipzh}.

\subsubsection {SU(3) relations between $D^+$ and $D_s$ decays}

Both of the following ratios of branching ratios 
 \beq{QQ1a}     
{{BR( D_s \rightarrow  K^+K^+\pi^-)}\over{BR(D_s \rightarrow  
K^+K^-\pi^+)}} \approx O(tan^4 \theta_c)   
\eeq 
\beq{QQ1b}     
{{BR( D^+ \rightarrow  K^+\pi^+\pi^-)}\over{BR(D^+\rightarrow  
K^-\pi^+\pi^+) }}  
\approx O(tan^4 \theta_c)   
\eeq
are ratios of a doubly Cabibbo forbidden decay to an allowed
decay and should be of order $tan^4 \theta_c$. The SU(3) transformation $d
\leftrightarrow s$ takes the two ratios (\ref{QQ1a}) and (\ref{QQ1b}) into the
reciprocals of one another. SU(3) requires
the product of these two ratios to be EXACTLY $\tan^8 \theta_c$\cite{plipzh}.
 \bea
tan^8 \theta_c = ~ ~ ~ ~ ~ ~ ~ ~ ~ ~ ~ ~ ~ ~ ~ ~ ~ ~ ~ ~ ~ ~ 
~ ~ ~ ~ ~ ~ ~ ~ ~ ~ ~~ ~ ~ ~ ~ ~ ~ ~ ~ ~ ~ \nonumber \\
{{BR( D_s \rightarrow  K^+K^+\pi^-)}\over{BR(D_s \rightarrow K^+K^-\pi^+)}}
\cdot {{BR( D^+ \rightarrow  K^+\pi^+\pi^-)}\over{BR(D^+\rightarrow
K^-\pi^+\pi^+) }} 
\eea 

Most  obvious
SU(3)-symmetry-breaking factors cancel out in this product; e.g. phase space.
Present data\cite{PDG} show
\beq{dub1}
 {{BR(D^+\rightarrow K^+\pi^-\pi^+)}\over{BR(D^+ \rightarrow K^-\pi^+\pi^+)}}
  \approx
  0.65\% \approx 3 \times \ tan^4 \theta_c
\eeq
Then SU(3) predicts
\beq{dub2}
 {{BR(D_s\rightarrow K^+K^+\pi^-)}\over{BR(D_s\rightarrow K^+K^-\pi^+)}}
\approx
{{tan^4 \theta_c}\over{3}} \approx  0.07\%.
\eeq  

If this SU(3) prediction is confirmed
experimentally some new dynamical explanation will be needed for the 
order of magnitude difference between effects of the final-state
interactions in $D^+$ and $D_s$ decays. 

If the final state interactions  behave similarly 
in $D_s$ and $D^+$ decays,
the large violation of SU(3) will need some explanation.

New physics enhancing the doubly suppressed decays
might produce a CP violation  observable as a charge asymmetry in the products
of above the two ratios; i.e between the values for $D^+$ and $D_s$ decays and
for $D^-$ and $\bar D_s$ decays.

An obvious caveat is the almost trivial SU(3) breaking arising
from resonances in the final states. But  sufficient data and Dalitz plots
should enable including these effects.
In any case the SU(3) relation and its possible violations raise interesting
questions which deserve further theoretical and experimental investigation.
Any really large SU(3)-breaking final state interactions
that we don't understand must cast serious doubts on many SU(3)
predictions. 

\subsubsection {A problem with strong phases}.
 
The $d \leftrightarrow s$ interchange SU(3) 
transformation also predicts\cite{lincoln}
  $ D^o \rightarrow K^+ \pi^-$ and $ D^o \rightarrow K^- \pi^+ $  
have the same strong phases. 
This has been shown 
to be in disagreement with experiment\cite{sven} showing SU(3) violation.  
   
But the $ K^+
\pi^- $ and $ K^- \pi^+ $ final states are charge conjugates of one another
and strong interactions conserve charge conjugation.
SU(3) can be
broken in strong interactions without breaking charge conjugation only in the
quark - hadron form factors arising in hadronization transitions like  
 \bea 
D^o(c\bar u) &
\rightarrow  (s \bar u\rightarrow K^{-} )_S
\cdot  (u\bar d\rightarrow  a_1^+ )_W  \nonumber \\ 
& \rightarrow  K^{-} a_1^+  \rightarrow
K^-\pi^+ 
\eea 
 \bea D^o(c\bar u) & 
\rightarrow  (d \bar u \rightarrow a_1^-)_S \cdot (u\bar s \rightarrow K^{*+} )_W
 \nonumber \\ 
&\rightarrow  a_1^-  K^{+}  \rightarrow \pi^- K^+ 
\eea  
with the SU(3)  breaking given by the inequality (19).

The $a_1$ and $\pi$  wave functions are very different and  not related by
SU(3). 
The $ K^{\mp} a_1^\pm \rightarrow K^\mp\pi^\pm$ transition
can proceed via $\rho$ exchange 

\section{Puzzles and Challenges in B Decays }

Weak Decays need hadron models and QCD to interpret decays, but have 	
too many diagrams and too many free parameters.  
Use of flavor topology can simplify analyses on one hand and
challenge QCD to explain them if they work.

\subsection{{OZI in Heavy Flavor Decays}}

Two flavor topology predictions which challenge conventional
wisdom\cite{bkpfsi};   
\beq{nunudnika}
BR (B^\pm \rightarrow K^\pm \omega) = BR (B^\pm \rightarrow K^\pm \rho^o)
\eeq
\beq{nunudnkb}
 \tilde \Gamma(B^\pm \rightarrow K^\pm \phi)
= \tilde \Gamma(B^\pm \rightarrow K^o \rho^\pm)
\eeq
where $\tilde \Gamma$ denotes the predicted partial width when phase space
differences are neglected.The first (\ref{nunudnika}) assumes only the
exclusion of ``hairpin diagrams" and holds even in presence of strong
final state rescattering via all other quark-gluon diagrams. The second
(\ref{nunudnkb}) also assumes SU(3) flavor symmetry between strange and
nonstrange pair production 

\subsection{Anomalously high $\eta'$ in many final states} 

The large experimental branching ratio\cite{PDG} 
$BR(B^+ \rightarrow  K^+ \eta')=6.5\pm 1.7 \times 10^{-5}$ as compared with
$BR(B^+ \rightarrow  K^+ \eta) < 1.4 \times 10^{-5}$ 
and $BR(B^+ \rightarrow  K^o \pi^+)=2.3\pm 1.1 \times 10^{-5}$ 
still has no completely satisfactory explanation and has aroused considerable
controversy\cite{PHAWAII,Lipkin2000} Also the large inclusive 
$B^+ \rightarrow  K^+ \eta' X$ branching ratio is equally puzzling.

A parity selection rule can separate the two types of models proposed to
explain this  high $\eta'$ appearance  and the high $\eta'/\eta$ ratio.

1. The OZI-forbidden hairpin diagram\cite{bkpfsi} predicts a  universal
$parity$-independent enhancement for all  final states arising from the flavor
singlet component of the $\eta'$ \cite{Atwood:1997bn,Halperin:1998ma}.

2.  Parity-dependent interference
between diagrams producing the $\eta'$ via its strange and nonstrange 
components\cite{bkpfsi} predicts a large 
$\eta'$/$\eta$ ratio for even parity final states like $K\eta$ and
$K\eta'$  the reverse for  odd parity states like
$K^*$(892) $\eta$ and $K^*$ $\eta'$\cite{PHAWAII}.
So far this selection rule agrees with experiment.


\begin{thebibliography}{9}

\bibitem{natmil} Harry J. Lipkin, Nature 406 (2000) 127

\bibitem{hiller} Gudrun Hiller, hep-ph/0207356
 
\bibitem{kagneub} Thomas Becher et al,,  hep-ph/0205274 Phys. Lett. B540
(2002) 278

\bibitem{nucabib}  M. Karliner and H.J. Lipkin, hep-ph/0202099 .Phys. Lett. B533
(2002) 60

\bibitem{CCFR}
CCFR Collab.
A.O.~Bazarko {\it et al.},
%``Determination of the strange quark content of the nucleon from a next-to-
%                 leading order QCD analysis of neutrino charm production,"
Z. Phys. {\bf C65}, 189 (1995);

\bibitem{Iswis}{N. Isgur and M. B. Wise, Phys. Rev. Lett. 66 (1991) 1130}

\bibitem{Ma} D. Ashery and H.J. Lipkin, Phys.Lett. B469 (1999) 263
 hep-ph/9908355 and hep-ph/0002144 
 
\bibitem{PDG}{Particle Data Group, Eur. Phys. J. C 15 (2000) 1}

\bibitem{vecdom} Harry J. Lipkin,  
%Systematics of Large Axial Vector Meson
%Production in Heavy Flavor Weak Decays 
%Tel Aviv University preprint TAUP 2655-2000
%Weizmann Preprint WIS/26/00-DEC.-DPP
%Argonne Preprint ANL-HEP-PR-00-126
hep-ph/0011228, 
Physics Letters B 515 (2001) 81 

\bibitem{nuclolip}F. E. Close and H. J. Lipkin, hep-ph/0208217

\bibitem{Lipkin1997}
H.~J.~Lipkin,
%``Penguins, trees and final state interactions in B decays in broken  SU(3),''
Phys.\ Lett.\ { B415}, 186 (1997)
[hep-ph/9710342].

\bibitem{plipzh} Harry J. Lipkin and Zhi-zhong Xing, Phys. Lett. B450 (1999)
405 

\bibitem{lincoln} L. Wolfenstein, Phys. Rev. Lett. 75 (1995) 2460

\bibitem{sven} S. Bergmann et al,
Phys.Lett. B486 (2000) 418; hep - ph/0005181 

\bibitem{bkpfsi}H.~J.~Lipkin,
Phys.\ Lett.\ { B433}, 117 (1998)

\bibitem{PHAWAII}Harry J. Lipkin,
In Proceedings of the 2nd International Conference on B
      Physics and CP Violation, Honolulu, Hawaii, 24-27 March 1997
Edited by T. E. Browder et al, World Scientific, (1998) p.436
hep-ph/9708253


\bibitem{Lipkin2000} See for example
H.~J.~Lipkin,
Phys.\ Lett.\ { B494}, 248 (2000)
[hep-ph/0009241].
%%CITATION = HEP-PH 0009241;%% and
A.~Datta, X.~G.~He and S.~Pakvasa,
%``Quasi-inclusive and exclusive decays of B to eta',''
Phys.\ Lett.\ { B419}, 369 (1998)
[hep-ph/9707259].



\bibitem{Atwood:1997bn}
D.~Atwood and A.~Soni,
%``B --> eta' + X and the QCD anomaly,''
Phys.\ Lett.\ { B405}, 150 (1997)
[hep-ph/9704357].
\bibitem{Halperin:1998ma}
I.~Halperin and A.~Zhitnitsky,
%``Why is the B --> eta' X decay width so large?,''
Phys.\ Rev.\ Lett.\ { 80}, 438 (1998)
[hep-ph/9705251].


\end{thebibliography}
\end{document}